# Simulation of the energy efficiency auction prices in Brazil

Javier L. L. Gonzales[1,2], Rodrigo F. Calili[2], Reinaldo C. Souza[2,3] and Felipe L. Coelho da Silva [3,4]

[1]E.P. de Ingeniería Ambiental
Universidad Peruana Unión (UPeU)
Lima, Perú
**Phone/Fax number: + (51) 980578889, e-mail: linkolklg@upeu.edu.pe**

[2]Department of Metrology
Pontifical Catholic University of Rio de Janeiro (PUC-Rio)
Rio de Janeiro, Brasil
Phone/Fax number: + (55 21) 3527 1542/ + (55 21) 3527-2060, e-mail: linkolklg@upeu.edu.pe, rcalili@esp.puc-rio.br, reinaldo@ele.puc-rio.br

[3]Department of Electrical Engineering
Pontifical Catholic University of Rio de Janeiro (PUC-Rio)
Rio de Janeiro, Brasil
Phone/Fax number:+ (55 21) 3527-1001, e-mail:reinaldo@ele.puc-rio.br, felipeleite@ufrrj.br

[4]Department of Mathematics
Federal Rural University of Rio de Janeiro (UFRRJ)
Rio de Janeiro, Brasil
Phone/Fax number:+ (55 21) 2682-1120, e-mail:felipeleite@ufrrj.br

**Abstract.** The electricity consumption behavior in Brazil has been extensively investigated over the years due to financial and social problems. In this context, it is important to simulate the energy prices of the energy efficiency auctions in the Brazilian regulated environment. This paper presents an approach to generate samples of auction energy prices in energy efficiency market, using Markov chain Monte Carlo method, through the Metropolis-Hastings algorithm. The obtained results show that this approach can be used to generate energy price samples.

## 1. Introduction

The behavior of electricity consumption in Brazil has been extensively investigated over the past years due to the economic and social importance that electricity has to the development of the country. Possible errors in the electrical sector planning and possible power supply problems can cause immense harm to the country. Thus, energy efficiency is very important to the decision-making bodies and entities that operate the energy sector.

The energy efficiency is considered a synonymous of environmental preservation, because the energy saved prevents the construction of new generating plants and transmission lines. According to the International Energy Efficiency Scorecard (ACEEE), the countries that lead the world ranking of energy efficiency include Australia, Brazil, Canada, China, France, Germany, India, Italy, Japan, Mexico, Russia, South Korea, Spain, the United Kingdom, the United States, and the European Union. There are 31 energy efficiency indicators that the ACEEE considered for the evaluation of different countries that include in their government policies, many energy efficiency mechanisms. Brazil is in the raking, but has a very low level of energy efficiency compared to leading countries such as Germany and Italy [1]. The Demand-Side Bidding could represent a very interesting alternative for the revitalization and promotion of energy efficiency practices in Brazil. However, it is important to note that this implies in reliable estimation of the amount of reduced energy, which cannot be carried out without the implementation and development of a measurement and verification system.

In Brazil, the energy auctions are processes bidding conducted in order to hire the electricity necessary to ensure full compliance of future demand in the regulated contracting environment, where the Distribution Utilities are inserted. In this context, this paper presents an approach to simulate the energy prices of the energy efficiency auction in the Brazilian regulated environment. The trading data from energy efficiency auction prices were obtained from previous auctions held in Brazil from 2005 to 2014. Those data were transformed using a factor to incorporate inflation. The samples of auction energy prices in energy efficiency were obtained using the method of Markov chain Monte Carlo, through the Metropolis-Hastings algorithm. Also, as the distribution of energy prices is unknown it was used the nonparametric method of Kernel to obtain an estimate of the density, and thus approach a polynomial of large enough order to estimate this density. It was generated samples with the following sizes: 500, 1000, 5000 and 10000. From the samples generated a study was conducted to check if they recovered the distribution of the original sample information, and also, if the probability of the prices stay within the range of R$ 110 and R$ 140 (prices proposed in the energy efficiency auction, according [2]).

## 2. Demand Side Bidding in Brazil Proposal

In this section, it is presented a proposal for the DSB model adapted to the Brazilian reality.

The DSB allows an active participation of energy consumers (demand) in the processes of pricing, maintaining the quality of energy supply. The demand side management has important implications for the overall efficiency of electricity supply, both on the economic side as well as on the environmental side. Also, in the long range it also implies in postponing or even avoiding the need for costly investments in power plants.

Besides that, it also may imply in reducing both; the size of the transmission lines and the construction of large hydropower plants. This is rather important, as the total costs could reduce considerably.

In most cases, the large generators produce high levels of $CO_2$ emissions, which could be reduced with the DSB mechanism.

Before presenting the proposed model for implementation of the DSB mechanism, it is shown in Figure 1 the drivers that impact this market.

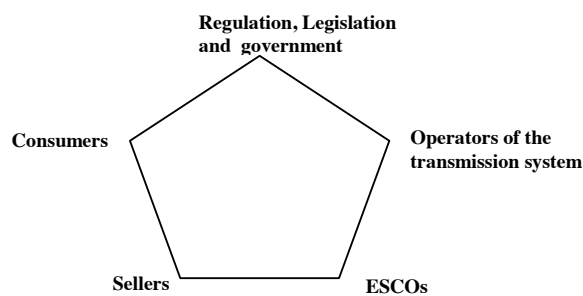

Figure 1: Actors in DSB mechanism (source [2]).

The DSB mechanism will not happen unless the different participants mutually benefit from them. The National Electric System Operator (ONS) is the responsible for maintaining the system security and the quality of the energy supply. And, the virtual power generation would be another energy power source available to ensure the system reliability.

Generally, in competitive markets, providers should buy enough energy to meet their customers' needs at all times, avoid possible financial penalties.

Customers have greater flexibility in their electricity usage patterns, being particularly valuable to providers, as they can help them avoid periods of peak energy demand.

Electricity consumers are naturally interested in buying electricity at the lowest cost, and management on the demand side can help in this case. Indeed, some consumers may change their consumer habits without any serious harm to their way of living. Moreover, in most cases, residential consumers use low amount of power, being not eligible to participate in the DSB mechanism.

Having examined the actors to participate in DSB mechanism, it is defined the concepts based on expert opinion [2]. Additionally, the steps include the attempt to adapt and use the technology to facilitate some points.

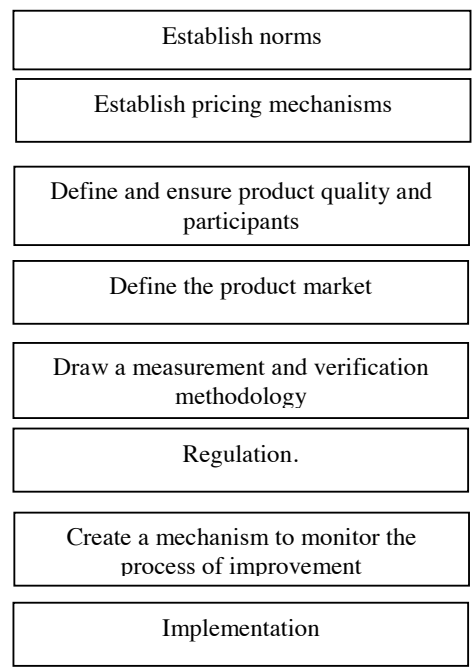

Figure 2: Demand Side Bidding proposal (source [2]).

The first step in developing a LEE system is to establish market governance standards/process. Moreover, the pricing mechanisms should be well defined, since before forming the price and set its sales policies, a market must have knowledge of the elements of formation of the product sales price.

Currently, many markets seek excellence in its production processes through quality, in the same sense, DSB mechanism tries to involve the definition and quality assurance of the participants.

Also, it is important to have a demand side product well adjusted. Linked to this, the design of a methodology for Measurement & Verification (M&V) will transmit reliability among its participants. In this context, it is essential to mention the regulation as it becomes an instrument by which the regulator can rely to establish rules (rights and obligations) to all participants. Beyond the regulator, Calili (2013) mentions that a mechanism to monitor the overall process should be created.

It is essential that: the efficiency and effectiveness are evaluated; the alternatives for problem solutions are explored; the best solution is selected and implemented.

## 3. Methodology

With the aid of the proposed methodology, are created the conditions for planners and controllers to better manage the DSB prices. Depending on how the price behaves over time, the proposed methodology provides the conditions to adapt. The methodology is presented in what follows. First, it is described a method for approaching a proper density function to the behavior of the original data (Energy Efficiency Auction prices). Then the Markov Chain Monte Carlo (MCMC) method, which generates samples from the distribution of interest, is presented. The MCMC algorithm used was the Metropolis-Hastings, which allows different samples to analyze different price scenarios.

### A. Kernel method and polynomial approximation

Data for Energy Efficiency Auctions of price negotiations are taken from previous auctions, from 2005 to 2014. The objective of this research was to evaluate the behavior of energy prices between R$ 110/MWh and R$ 140/MWh, as suggested by [2].

The data distribution in the range of interest is unknown. Therefore, the nonparametric kernel method was used to obtain the estimated density function of the data energy prices. The method can be interpreted as a generalization of the histogram. Then an estimated density function provided by the kernel method was approximated by a polynomial of order 17. The polynomial coefficients were obtained using the method of least squares.

### B. MCMC

Considering the estimated density obtained by the kernel method and polynomial approximation, one has the presumed distribution of interest and the MCMC method for generating samples can be used.

MCMC method can generate samples from a distribution of any interest. MCMC algorithm used was the Metropolis-Hastings (MH) with distribution of interest being obtained in the polynomial approximation. In fact, it was generated values of the random variable energy prices through the probability distribution obtained by the Kernel method.

### C. Algorithm Metropolis

The Metropolis-Hastings algorithm (MH) was initially developed by Metropolis (1953) and generalized by [3]. Indeed, the idea of the algorithm is to obtain the desired distribution, through the proposed distribution. In this case, the distribution data which is used to generate an accomplishment depends on the previous simulation. The distribution proposal adopted in this study was the uniform distribution.

The algorithm was able to do simulations and create scenarios with different sample sizes. The generated samples follow the behavior of the original data. Thus, one can use these samples to understand the behavior of the distribution of energy prices. And so, these samples have enabled the probability distribution of the proposed price of DSB between R$ 110/MWh and R$ 140/MWh, as suggested by Calili (2013).

## 4. Results

Through the nonparametric kernel method, it was possible to adjust a density of an order 17 polynomial function (Figure3).

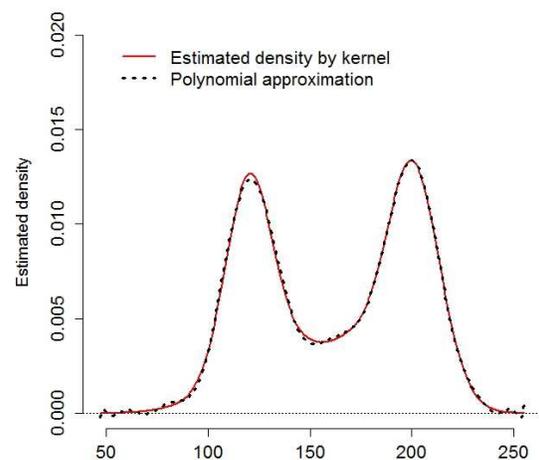

Figure 3: Density estimated by Kernel algorithm

With the achieved estimation, it is possible to generate simulations following the order of the polynomial. Thus, several samples were generated with different size. It can be considered the larger the sample the better is the approximation to the original data. In this respect, it is worth mentioning that both the use of kernel method and the use of MCMC are crucial to provide approximations of the original data and clearly analyze its impact. Next, it is observed the behavior of the data in each histogram with 500, 1,000, 5,000 and 10,000samples, considering only one scenario.

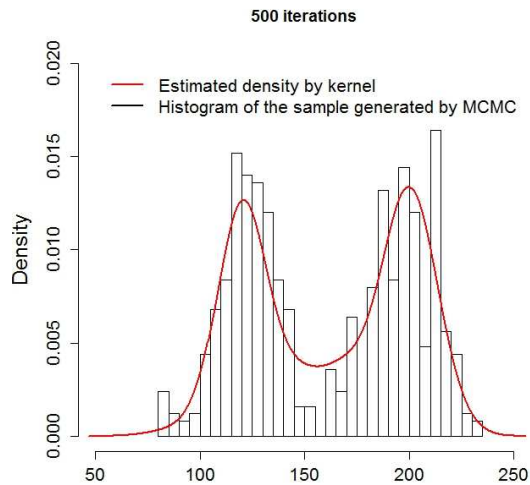

Figure 4: Histogram with Kernel and MCMC for 500 iterations

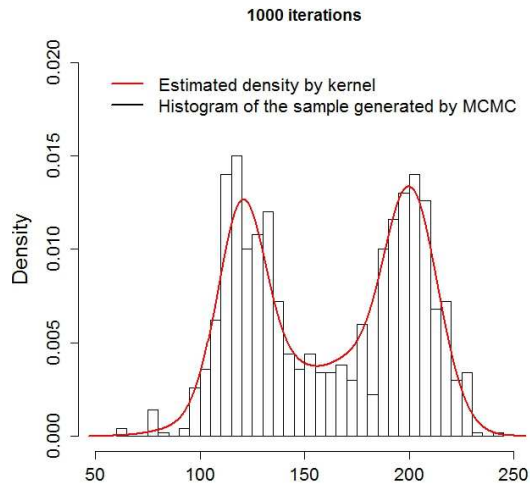

Figure 5: Histogram with Kernel and MCMC for 1,000 iterations

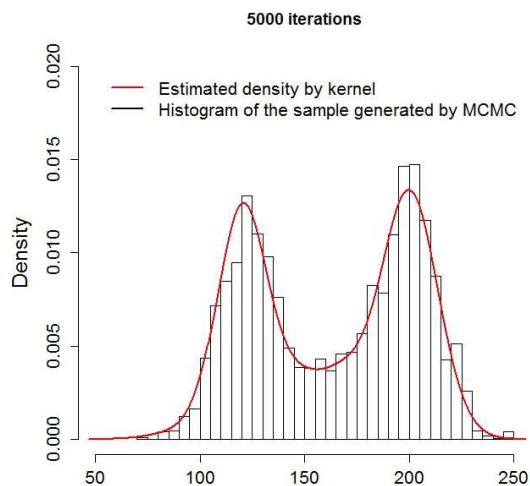

Figure 6: Histogram with Kernel and MCMC for 5,000 iterations

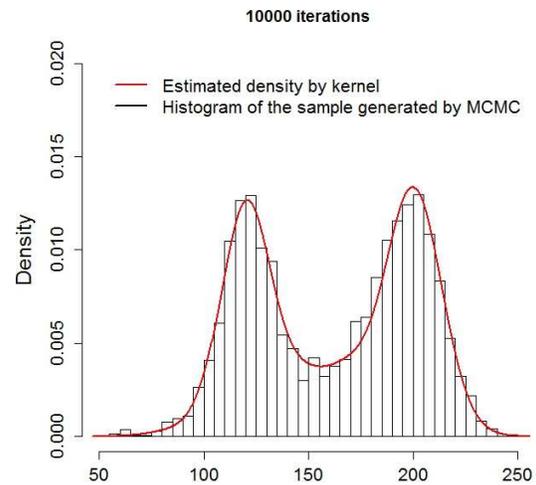

Figure7: Histogram with Kernel and MCMC for 10,000 iterations

It can be noted that the sample which best approximates the original data in accordance with the generated histograms, is the 10,000 size, having consistently followed the behavior of the data.

Moreover, once achieved the best-fitted curve, it was carried out the analysis of each scenario with each sample to find the probability of prices staying within the range R$110/MWh and R$ 140/MWh. The following table presents the different results obtained with each probability.

Table1: Probabilityin the Scenario 1

| Size of sample | Probability |
| --- | --- |
| 500 | 27.80% |
| 1,000 | 26.10% |
| 5,000 | 29.66% |
| 10,000 | 30.86% |

As shown in Table 1 in the scenario 1, the price falls in the range of R$ 110 and R$ 140 in 27.80% of the 500 sample; in the 1,000 sample, the probability to be in this range is 26.10%; considering the 5,000 sample, the probability is 29.66%; and in the 10,000 sample, it is 30.86%.

Table2: Probabilityin the Scenario 2

| Size of sample | Probability |
| --- | --- |
| 500 | 28.60% |
| 1,000 | 31.40% |
| 5,000 | 30.00% |
| 10,000 | 31.61% |

As shown in Table 2 in the scenario 2, the price falls in the range of R$ 110 and R$ 140 in 28.60% of the 500 sample; in the 1,000 sample, the probability to be in this range is 31.40%; considering the 5,000 sample, the probability is 30%; and in the 10,000 sample, it is 31.61%.

Table3: Probabilityin the Scenario 3

| Size of sample | Probability |
| --- | --- |
| 500 | 31.80% |
| 1,000 | 29.10% |
| 5,000 | 30.94% |
| 10,000 | 30.07% |

As shown in Table3 in the scenario3, the price falls in the range of R$ 110 and R$ 140 in 31.80% of the 500 sample; in the1,000 sample, the probability to be in this range is 29.10%; considering the 5,000 sample, the probability is 30.94%; and in the 10,000 sample, it is 30.07%.

Table4: Probability in the Scenario 4

| Size of sample | Probability |
| --- | --- |
| 500 | 25.80% |
| 1,000 | 30.80% |
| 5,000 | 30.90% |
| 10,000 | 29.88% |

As shown in Table 4 in the scenario 4, the price falls in the range of R$ 110 and R$ 140 in 25.80% of the 500 sample; in the 1,000 sample, the probability to be in this range is 30.80%; considering the 5,000 sample, the probability is 30.90%; and in the 10,000 sample, it is 29.88%.

Table5: Probability in the Scenario 5

| Size of sample | Probability |
| --- | --- |
| 500 | 34.60% |
| 1,000 | 27.80% |
| 5,000 | 30.20% |
| 10,000 | 32.18% |

As shown in Table 5 in the scenario5, the price falls in the range of R$ 110 and R$ 140 in 34.60% of the 500 sample; in the 1,000 sample, the probability to be in this range is 27.80%; consideringthe5,000 sample, the probability is 30.20%; and in the10,000sample, it is 32.18%.

Table6: Probability in the Scenario 6

| Size of sample | Probability |
| --- | --- |
| 500 | 27.80% |
| 1,000 | 31.30% |
| 5,000 | 31.28% |
| 10,000 | 30.84% |

As shown in Table6 in the scenario 6, the price falls in the range of R$ 110 and R$ 140 in 27.80% of the 500 sample; in the 1,000 sample, the probability to be in this range is 31.30%; considering the 5,000 sample, the probability is31.28%; and in the 10,000 sample, it is 30.84%.

Table7: Probability in the Scenario 7

| Size of sample | Probability |
| --- | --- |
| 500 | 35.20% |
| 1,000 | 32.10% |
| 5,000 | 29.42% |
| 10,000 | 28.66% |

As shown in Table7 in the scenario 7, the price falls in the range of R$ 110 and R$ 140 in 35.20% of the 500 sample; in the 1,000 sample, the probability to be in this range is 32.10%; considering the 5,000 sample, the probability is 29.42%; and in the 10,000 sample, it is 28.66%.

Table8: Probability in the Scenario 8

| Size of sample | Probability |
| --- | --- |
| 500 | 28.40% |
| 1,000 | 30.20% |
| 5,000 | 28.60% |
| 10,000 | 28.50% |

As shown in Table 8 in the scenario 8, the price falls in the range of R$ 110 and R$ 140 in 28.40% of the 500 sample; in the1,000 sample, the probability to be in this range is 30.20%; considering the 5,000 sample, the probability is 28.60%; and in the 10,000 sample, it is 28.50%.

Table9: Probability in the Scenario 9

| Size of sample | Probability |
|---|---|
| 500 | 30.40% |
| 1,000 | 26.20% |
| 5,000 | 30.04% |
| 10,000 | 31.07% |

As shown in Table 9 in the scenario 9, price falls in the range of R$ 110 and R$ 140 in 30.40% of the 500 sample; in the 1,000 sample, the probability to be in this range is 26.20%; considering the 5,000 sample, the probability is 30.04%; and in the 10,000 sample, it is 31.07%.

Table10: Probability in the Scenario 10

| Size of sample | Probability |
|---|---|
| 500 | 28.20% |
| 1,000 | 33.00% |
| 5,000 | 29.96% |
| 10,000 | 32.36% |

As shown in Table10 in the scenario 10, price falls in the range of R$ 110 and R$ 140 in 28.20% of the 500 sample; in the1,000 sample, the probability to be in this range is 33.00%; considering the 5,000 sample, the probability is 29.96%; and in the 10,000 sample, it is 32.36%.

## 5. Conclusions

This research dealt with the simulation of energy prices in the Brazilin regulated contracting environment considering a possible DSB mechanism.

The simulation provided experimentation, considering policy proposals using statistical theory. It has taken a considerable simulation processing to obtain significant results in statistical terms. However, it was possible to adapt the energy prices of a possible power DSB to a curve over Kernel method (through a polynomial of order 17). Furthermore, it performed the simulation via Markov chain Monte Carlo, and using the Metropolis-Hastings algorithm generated samples to better analyze the behavior of the prices.

As further work, another methodology to obtain, the different scenarios that were generated in the regulated contracting environment can be considered as a possible alternative to the proposed approach in this paper.

## Acknowledgements

The authors would like to thank the colleagues from PUC-Rio for their valuable comments and suggestions, which improved the paper and the R&D program of the Brazilian Electricity Regulatory Agency (ANEEL) for the financial support (R&D project PD-7625-0003/2014). Finally, the authors would like to thank CNPq and CAPES, which are governmental agencies that have also given support to this research.